\documentclass[useAMS,eqsecnum,usenatbib,twocolumn]{mn2e}

\usepackage{graphicx}
\usepackage{dcolumn}
\usepackage{bm}
\usepackage{subfigure}
\usepackage[]{natbib}
\usepackage{color}
\bibpunct{(}{)}{,}{a}{}{,}

\title[Separation of spacetime and matter in polar oscillations of compact stars]{Separation of spacetime and matter in polar oscillations of compact stars}

\author[Y. J. Zhang, Q. Q. Zhu, J.~Wu, T. K. Chan and P.T.~Leung]{
Y. J. Zhang\thanks{Present address: Department of Physics,
University of California, San Diego, 9500 Gilman Drive, La Jolla,
CA 92093, USA.}, Q. Q. Zhu, J. Wu\thanks{Present address: Office
of University General Education, The Chinese University of Hong
Kong, Shatin, Hong Kong SAR, China.}, T. K. Chan$^\star$ and
P. T.~Leung\thanks{Email: ptleung@phy.cuhk.edu.hk} \\
Physics Department and Institute of Theoretical Physics, The Chinese University of Hong Kong, Shatin, Hong Kong SAR, China.}

\begin{document}

\date{\today}

\pagerange{\pageref{firstpage}--\pageref{lastpage}} \pubyear{2013}

\label{firstpage}

\maketitle
\begin{abstract}
The dynamics of polar oscillations of compact stars are
conventionally described by a set of coupled differential
equations, and the roles that spacetime and matter play in such
oscillations are consequently difficult to track. In the present
paper, we develop a scheme for polar oscillations in terms of one
spacetime and one matter variables. With our judicious choice of
variables, the coupling between spacetime and matter is weak,
which enables the separation of these two parties. Two independent
second-order ordinary differential equations are obtained,
corresponding to the Cowling and inverse-Cowling approximations,
and leading to accurate determination of {\it p} and {\it w}-mode
quasi-normal modes, respectively, in a unified framework.
\end{abstract}




\def\i{ {\rm i}}
\def\r{ {\rm r}}

\section{Introduction}
Compact stars, including neutron stars (NSs) and quark stars
(QSs), are notable for the high density achievable at their
centers and the huge gravitational field surrounding them. Their
internal structure  is often the ideal test bed for the equation
of state (EOS) of nuclear (or quark) matter and can provide clues
to the interactions between elementary particles as well
\citep[see, e.g.,][and references
therein]{ComStar,lattimer2007nso,Weber-pulsar,Lattimer_2012}.
Gravitational waves (GWs) emitted from oscillating NSs (or QSs)
are expected to carry rich information about their internal
structures and hence the physics of dense matter as well, giving
rise to various efforts to infer the properties of relativistic
compact stars from close examination of their oscillation spectra
\citep[see,
e.g.,][]{Andersson1996,Andersson1998,Ferrari,Kokkotas_2001,
Ferrari_prd,Tsui05:3,Tsui:prd,lau-2009,PhysRevD.83.024014,Kokkotas_2013PRD}.
Although the oscillation frequencies of NSs usually lie outside
the sensitivity window of the existing GW detectors such as
GEO600, LIGO I, TAMA300 and VIRGO \citep[see,
e.g.,][]{Hughes_03,Fryer:2001zw}, with the advent of more
sensitive third-generation GW telescopes \citep[e.g., the proposed
Einstein Telescope, see][]{Andersson:2009yt,Hild:2010id}, it is
probable that the structure and the composition of a compact star
can be inferred from its GW signals in the coming future.

The fully relativistic formulation of stellar pulsations was first
founded by \citet{Thorne-1967-1} and since then alternative
approaches to such processes have emerged in the literature
\citep[see,
e.g.,][]{Lindblom-1983,Lindblom-1985,Chandrasekhar1,Lindblom-1997,Kokkotas_rev}.
Basically, oscillations of compact stars are classified into axial
and polar types. For axial type oscillations, the matter forming a
star is essentially a spectator and only the spacetime surrounding
the star pulsates. For polar type oscillations, matter motion and
spacetime variation are, however, coupled together. To describe
such coupled motion, different forms of equations of motion have
been proposed since the pioneering work of \citet{Thorne-1967-1}.
For example, based on \citet{Thorne-1967-1}, \citet{Lindblom-1983}
and \citet{Lindblom-1985} described the system with four coupled
first-order ODEs in four variables (termed as the LD formalism in
this paper), which can be readily solved numerically. Later, to
draw an analogy between the oscillations of relativistic and
Newtonian stars, \citet{Lindblom-1997} proposed a new formalism to
describe the coupled oscillation with two second-order equations
(termed as the LMI formalism), which reduce directly into the
corresponding Newtonian equations under the weak field limit. In
another relativistic stellar pulsation formalism proposed by
\citet{Allen-1998} (termed as the AAKS formalism in this paper),
variations in matter and spacetime are described by one matter
variable and two metric variables. The three variables are
governed by three coupled second-order time-dependent wave
equations. \citet{Allen-1998} managed to evolve these three
variables numerically in the time domain for certain sets of
initial conditions and studied  the characteristics of the GWs
emitted in such cases.

Polar type oscillations are very much involved in dynamical
processes such as supernova and binary mergers of NSs. The
interplay between spacetime and matter in polar oscillations of
compact stars, however, often poses hurdles for developing a firm
grasp of these phenomena. To trace the individual roles played by
matter and spacetime components, different variants of the Cowling
approximation (CA), where effects of metric perturbations are
neglected \citep{MVSCowling,Finn1988,Lindblom-1990,
Lindblom-1997}, and the inverse-Cowling approximation (ICA), where
fluid motion is ignored \citep{Andersson-1996, Wu-2007}, have been
proposed. These approximations vary in the choice of the physical
variable measuring the matter or spacetime component and lead to
different degrees of accuracy (see the above-mentioned
references).

In the present paper, we study polar oscillations with
quasi-normal mode (QNM) analysis. QNMs are oscillation modes with
all dynamical variables decay exponentially in time
\citep{Press_1971,Leaver_1986,rmp,Kokkotas_rev,Berti:2009kk,RevModPhys.83.793}.
Mathematically, the perturbation variables are assumed to have the
$e^{-i\omega t}$ dependence , where the eigenfrequency $\omega
\equiv \omega_\r-i\omega_\i$ is complex-valued with $\omega_\i
\geq 0$ measuring the decay rate of the mode. QNMs of compact
stars reveal the physical characteristics (e.g., mass, radius,
moment of inertia, EOS and composition) of a star \citep[see,
e.g.,][]{Andersson1996,Andersson1998,Ferrari,Kokkotas_2001,
Ferrari_prd,Tsui05:3,Tsui:prd,lau-2009}. Polar QNMs of compact
stars can be classified into the fluid mode, including the
fundamental $f$-mode, the pressure $p$-mode and the gravity
$g$-mode according to the nature of respective restoring forces
\citep[][]{Cowling}, and the spacetime $w$-mode which has no
Newtonian counterpart.

The aim of the present paper is to propose a physically more
transparent picture for polar oscillations of compact stars. Two
coupled second-order differential equations in the frequency
domain, involving one matter field and one spacetime variable used
in the AAKS formalism, are derived (see Section 2 and Appendix
\ref{SI-A}). Our judicious choice of the two variables enables a
straightforward decoupling of the two equations, leading to two
independent equations accurately describing fluid modes
(especially $p$-modes) and the $w$-modes, respectively (see
Section 3). Hence, in a unified framework, we simultaneously
establish both CA and ICA, which can lead to much improved
accuracies in comparison with other approximation schemes proposed
previously (see Sections 4 and 5 respectively). Under the current
CA, fluid motion is governed by a Sturm-Liouville eigenvalue
system, where completeness and orthogonality relations hold as
usual \citep[see, e.g.,][]{Morse-2008}. Therefore, the fluid
system is a conservative one and can be studied with conventional
normal-mode analysis. The mode with the lowest eigenfrequency
indeed corresponds to the $f$-mode, while the following modes are
good approximations for the successive $p$-modes. Besides, we also
find that other variants of CA \citep{MVSCowling,Finn1988} can
also be recast into a form analogous to ours, save the difference
in two parameters. As regards the present ICA, to our knowledge,
this is the first time that an approximate single second-order
differential equation describing polar $w$-mode QNMs, which has
been numerically verified to lead accuracies much better than one
percent,  has been formulated.

Geometric units in which $G=c=1$ are adopted throughout this
paper. We consider polar oscillations of a non-rotating compact
star of perfect fluid with radius $R$ and total mass $M$. Unless
stated otherwise, all numerical values of frequencies are measured
in units of $M^{-1}$ and evaluated for the case where the angular
momentum index $l$ equals 2. Besides, we also assume that the
effect of temperature on the EOS of nuclear matter is negligible
and hence $g$-modes are omitted in our discussion. In order not to
divert the reader's attention from the main physical findings
achieved in the paper, we put most of the mathematical details in
three appendices.

\section{Two-variable scheme for Polar QNMs}
\subsection{Three-variable scheme}
When a static compact star is perturbed, its fluid and spacetime
acquire small-amplitude, time-dependent perturbations about the
equilibrium state. These perturbations are first decomposed into
tensorial and vectorial spherical harmonics $Y_{lm}(\theta, \psi)$
with angular momenta $\{lm\}$, and then separated into axial and
polar parts according to parity. The perturbed line element of
polar parity in the Regge-Wheeler gauge is
\citep{Thorne-1967-1,RWeq,Thorne-1969-2,Kojima92}:
\begin{eqnarray}
ds^2\!\!\!\!\!&=&\!\!\!\!\!-e^{\nu}(1-H_0Y_{lm})dt^2+ e^{\lambda}(1+H_0Y_{lm})dr^2\nonumber\\
&&\!\!\!\!\!+2H_1Y_{lm}dtdr
 +r^2(1+KY_{lm})(d\theta^2+\sin^2\theta
d\phi^2),\label{polar-metric-perturbation}
\end{eqnarray}
where the convention of the metric follows that used in
\citet{Kojima92}, and $e^{\nu}$ and $e^{\lambda}$ are
characteristics of the star at equilibrium \citep[see, e.g.,
Chapter 2,][]{ComStar}.

In the AAKS formalism \citep{Allen-1998}, polar oscillations of
compact stars are described by two spacetime metric variables $F$,
$S$ and one fluid perturbation variable $H$, which are related to
the above metric perturbations and the Eulerian change in pressure
$\delta {p}$ as follows:
\begin{eqnarray}
S\!\!\!\!\!&=&\!\!\!\!\! e^{\nu}(H_0-K)/r, \nonumber\\
F\!\!\!\!\!&=&\!\!\!\!\! r K, \nonumber\\
H\!\!\!\!\!&=&\!\!\!\!\! \delta p/(\rho+p).\label{SFH-define}
\end{eqnarray}
The three variables $S$, $F$ and $H$ satisfy three coupled
second-order differential equations and a Hamiltonian constraint
(see Appendix \ref{SI-A}).

It has been shown that inside the star and in the weak field limit $K$ and $H_0$ are related to the perturbation in the Newtonian potential $U_1$  by \citep{Thorne-1969-4}:
\begin{equation}\label{weak}
K \simeq H_0 \simeq 2U_1.
\end{equation}
Therefore, following directly from (\ref{SFH-define}), to leading
orders in field strengths, $S$ and $F$ measure the relativistic
and Newtonian effects of gravity, respectively. Hence, in order to
describe generation of GWs, which is the dominant process in the
spacetime mode, $S$ is the right variable to keep. On the other
hand, $H$ is proportional to the Eulerian change in pressure and
describes the motion of matter. We therefore eliminate the
variable $F$ in the AAKS formalism and establish a scheme with one
spacetime variable $S$ and one matter variable $H$.

\subsection{Two-variable scheme}
For QNMs, the three field variables  in AAKS formalism take the
 time-dependence $e^{-i\omega t}$, e.g., $S(r, t)=S(r)e^{-i\omega
t}$ and etc. A  two-variable scheme is achievable by eliminating
one of these three variables in the frequency domain.   In
particular, in the so-called $H$-$S$ scheme, the variable $F(r)$
and its derivative  are expressed in terms of $S(r)$, $H(r)$ and
their derivatives  (see Appendix \ref{SI-B}). Two coupled
second-order ODEs inside the star are then readily obtained:
\begin{eqnarray}
\frac{d^2 S}{d r_*^2}\!\!\!\!\!&+&\!\!\!\!\!U_{S1}(\omega, r)\frac{d S}{dr_*} +[\omega^2-V_{S0}(r)-U_{S0}(\omega, r)]S\nonumber\\\!\!\!\!\!&=&\!\!\!\!\!\Delta_{H1}(\omega, r)\frac{d H}{d r_*}+\Delta_{H0}(\omega, r)H, \label{AAKS-S-inside}\\
\frac{d^2 H}{d r_*^2}\!\!\!\!\!&+&\!\!\!\!\![V_{H1}(r)+U_{H1}(\omega, r)]\frac{d H}{dr_*}+[\frac{\omega^2}{C_s^2}-V_{H0}(r)\nonumber\\\!\!\!\!\!&-&\!\!\!\!\!U_{H0}(\omega, r)]H=\Delta_{S1}(\omega, r)\frac{d S}{d r_*}+\Delta_{S0}(\omega, r)S. \label{AAKS-H-inside}
\end{eqnarray}
where
\begin{eqnarray}
V_{S0}(r) \!\!\!\!\!&=&\!\!\!\!\!\frac{2e^{\nu}}{r^3}\left[(n+1)r-2\pi r^3(\rho+3p)-m\right], \label{V-S0}\nonumber\\
V_{H0}(r) \!\!\!\!\!&=&\!\!\!\!\!\frac{2e^{\nu}}{r^2}\left[n+1-2\pi r^2(\rho+p)(3+\frac{1}{C_s^2})\right], \label{V-H0}\nonumber\\
V_{H1}(r)
\!\!\!\!\!&=&\!\!\!\!\!\frac{e^{(\nu+\lambda)/2}}{r^2}\left[(m+4\pi
pr^3)(1-\frac{1}{C_s^2})+2(r-2m)\right], \label{V-H1}\nonumber
\end{eqnarray}
with $ n=(l+2)(l-1)/2$, the tortoise coordinate $r_*(r) = \int_0^r
e^{[\lambda(r')-\nu(r')]/2}dr'$, and $C_s(r) \equiv (d p/d
\rho)^{1/2}$ being  the sound speed in the stellar fluid. The
expressions of  $U_{i}$ and $\Delta_{i}$, where $i=S0, S1, H0,
H1$, can be found in Appendix \ref{SI-B}. Their contributions are
typically much smaller than those of the $V$ terms. While
$\Delta_{i}$ measure the interaction between $H$ and $S$, $U_{i}$
represent the effects of $F$ on the other two fields.

Outside the star, $H$ is identically zero and the system is
described by only one second-order ODE in $S$:
\begin{eqnarray}
\frac{d^2 S}{d r_*^2}+U_{S1}(\omega, r)\frac{d S}{dr_*} +[\omega^2-V_{S0}(r)-U_{S0}(\omega, r)]S=0,\label{AAKS-S-out}
\end{eqnarray}
where all potential terms  appearing in (\ref{AAKS-S-out}) can be
obtained from their corresponding terms inside the star in
(\ref{AAKS-S-inside}) by setting $m(r)=M$, $\rho=p=0$.

\subsection{Evaluation of QNMs}
QNMs of vibrating compact stars are defined by the physical
solutions of the proposed $S$ and $H$ equations with the spacetime
variable $S$ satisfying the outgoing boundary condition $S
\rightarrow \exp(i\omega r_*$) at spatial infinity. In solving the
$S$-$H$ equation set inside the star, we numerically integrate
(\ref{AAKS-S-inside}) and (\ref{AAKS-H-inside}) outward to find
$S$ and $H$. First of all,  the regularity boundary condition for
$S$ and $H$ at $r=0$ leads to the asymptotic behavior $S \approx
s_0r_*^{l+1}$ and  $H \approx  h_0r_*^l$  near the origin, where
$s_0$ and $h_0$ are constants. By keeping only terms of order
$1/C_s^2$ in (\ref{Allen-H-w}) as $C_s \rightarrow 0$ when
approaching the stellar surface,  one can show that
\citep{Allen-1998,Ruoff}:
\begin{eqnarray}
&&-\omega^2{H}+\frac{M}{R^2}\frac{d H}{d
r_*}-\frac{M}{2R^3}\frac{d F}{d
r_*}+\frac{M}{2(R-2M)}\frac{d S}{d r_*} \nonumber \\
&&+\frac{M(R+2M)F}{2R^5}+\frac{MS}{2R(R-2M)} = 0.
\label{surface-H}
\end{eqnarray}
Expressing $F$ and its derivative in terms of the other two
variables, we can then impose the above boundary condition at the
stellar surface
 to fix the ratio of $s_0$ to $h_0$ to obtain a unique (up to some multiplicative constant) solution
of $S$ and $H$ inside the star. The inside and outside $S$
function is then connected at the star surface to determine the
QNM frequencies.

 In the following discussion, we
will show that the $H$-$S$ approach achieved here naturally forms
the basis of a decoupling scheme, and readily leads to accurate CA
and ICA for $p$-modes and $w$-modes, respectively.

\section{Separation of Spacetime and Matter}\label{Decouple}
Dynamical processes such as supernova and binary mergers of NSs
involve both spacetime variation and matter motion. Likewise, when
a  NS is perturbed away from equilibrium, the excitation energy is
initially stored partially in the spacetime and partially in the
matter. The energy stored in spacetime is damped out rapidly
through the emission of $w$-mode gravitational waves owing to the
short relaxation time (large imaginary part of eigenfrequency) of
$w$-modes. On the other hand, the energy in matter has to be
transferred into the spacetime through the coupling between matter
and spacetime, which is measured by the $\Delta_i$ terms in
(\ref{AAKS-S-inside}) and (\ref{AAKS-H-inside}), and is then
brought away by fluid mode gravitational waves. The damping of
fluid modes are usually slow due to the smallness of $\Delta_i$,
leading to   small imaginary parts of their eigenfrequencies,
especially for $p$-modes.

Our choice of spacetime and matter variables provide a direct view
on the weak coupling  between spacetime and fluid mode
oscillations. We therefore conclude that in fluid modes
(especially the $p$-modes), the associated spacetime oscillations
and hence GW emissions are almost negligible, whereas in the
spacetime $w$-modes, fluid oscillations are hardly involved. This
strongly suggests that the spacetime and matter in the developed
schemes can be decoupled by taking the approximation $\Delta_i=0$
in (\ref{AAKS-S-inside}) and (\ref{AAKS-H-inside}):
\begin{eqnarray}
\frac{d^2 S}{d r_*^2}\!\!\!\!\!&+&\!\!\!\!\!U_{S1}(\omega, r)\frac{d S}{dr_*} +[\omega^2-V_{S0}(r)-U_{S0}(\omega, r)]S=0, \label{AAKS-S-decoupled}\\
\frac{d^2 H}{d r_*^2}\!\!\!\!\!&+&\!\!\!\!\![V_{H1}(r)+U_{H1}(\omega, r)]
\frac{d H}{dr_*}+[\frac{\omega^2}{C_s^2}-V_{H0}(r)\nonumber\\\!\!\!\!\!&-&\!\!\!\!\!U_{H0}(\omega, r)]H=0. \label{AAKS-H-decoupled}
\end{eqnarray}

\section{Cowling Approximation for Fluid Modes} \label{CA-p-mode}
It is well known that the eigenfrequencies of the fluid modes,
including both $f$ and $p$-modes, of a compact star are
characterized by very small imaginary parts. The smallness of the
imaginary part implies the tininess of GWs emitted in these modes,
due to their weak couplings to spacetime variables. The decoupled
equation for the fluid variable $H$ is given by
(\ref{AAKS-H-decoupled}) in Section~\ref{Decouple}. In the
following, we further simplify it and develop a CA which is highly
accurate for $p$-modes, and also works reasonably well for
$f$-mode.


\subsection{Fluid equation}

The potentials $U_{H0}$ and $U_{H1}$ in (\ref{AAKS-H-decoupled})
have a common denominator $D(\omega,r)$ (see Appendix \ref{SI-B}),
\begin{eqnarray}
D\!\!\!\!\!&=&\!\!\!\!\! \omega ^4+\omega^2\frac{e^{\nu}}{r^4}\left[e^{\lambda}(m+4 \pi r^3 p)^2+4 m r+4 \pi r^4 (3 p+\rho)\right]\nonumber\\
\!\!\!\!\!&+&\!\!\!\!\! \frac{2 e^{2\nu}}{r^6}\left\{8 \pi r^3 (m+2 \pi r^3 p)(p+\rho)- e^{\lambda}(m+4 \pi r^3 p)^2\right. \nonumber\\
\!\!\!\!\!&\times&\!\!\!\!\! \left.\left[1+n-8 \pi r^2
(p+\rho)\right]\right\}, \label{D}
\end{eqnarray}
which can drop to zero at a certain $r$ at low frequencies,
typical of $f$-mode. On the other hand, $D(\omega,r)$ increases as
$\omega^4$ at high frequencies, typical of $p$-modes. We find that
for $p$-mode oscillations the effects of $U_{H0}$ and $U_{H1}$ are
pretty small (both terms decreases as $1/\omega^2$), and further
simplify (\ref{AAKS-H-decoupled}) into:
\begin{eqnarray}
\frac{d^2 H}{dr_*^2}+V_{H1}(r)\frac{d
H}{dr_*}+\left[\frac{\omega^2}{C_s^2}-V_{H0}(r)\right]H=0.\label{Cowling-H}
\end{eqnarray}
The above equation can also be derived by simply neglecting the
$S$, $dS/dr_*$, $F$ and $dF/dr_*$ terms in $H$ equation in AAKS
formalism \citep[see][and also Appendix A]{Allen-1998}. Unlike
$U_{H0}$ and $U_{H1}$, which are frequency-dependent, $V_{H0}$ and
$V_{H1}$ are both frequency-independent. The mathematical
structure of (\ref{Cowling-H}) is more desirable than
(\ref{AAKS-H-decoupled}) and leads to a complete orthonormal set
of eigenfunctions. Eq.~(\ref{AAKS-H-decoupled}), on the other
hand, properly accounts for the effects of $F$ on the fluid motion
and is found to be the best available approximation for $p$-modes
in the following discussion.

\subsection{Boundary conditions}
Under CA (\ref{Cowling-H}),  $p$-mode oscillations are determined
from (\ref{Cowling-H}) by imposing proper boundary conditions on
$H$ at $r=0$ and $r=R$.  Near the star center, $H \approx h_0
r_*^l$ as mentioned previously. Assuming that the EOS acquires the
polytropic form $p \propto \rho^{1+1/N}$ with the polytropic index
$N>1$ near the star surface, which is typical for realistic NSs,
we find that $C_s^2 \sim A_c (R_*-r_*)$ with $A_c=M/(R^2N)$ and
$R_*$ being the tortoise radius of the star. Consequently, the
$H$-equation near the stellar surface reduces asymptotically to
\begin{eqnarray}
\frac{d^2 H}{dr_*^2}-\frac{M}{R^2 A_c (R_*-r_*)}\frac{d
H}{dr_*}+\frac{\omega^2}{A_c (R_*-r_*)}H=0. \nonumber
\end{eqnarray}
There are two independent solutions for this equation. The first
one is bounded and behaves as:
\begin{eqnarray}
H \propto \left[1 + \omega^2 R^2
(r_*-R_*)/M\right].\label{H-surface}
\end{eqnarray}
The other one is proportional to $(R_*-r_*)^{1-N}$ and is
unbounded. As $H$ is a physical quantity and must be finite at the
stellar surface, it therefore follows (\ref{H-surface}).

The single $H$-equation (\ref{Cowling-H}) together with these two
boundary conditions form an eigenvalue problem. The approximate
eigenfrequencies of fluid modes under the CA can then be located.

\subsection{Completeness and orthogonality}

The eigenvalue problem of the purely matter-based $H$-equation (\ref{Cowling-H}) is actually a Sturm-Liouville eigenvalue problem, as it can be cast into the following standard form:
\begin{eqnarray}
\frac{d}{dr_*}\left[P(r_*)\frac{d
H}{dr_*}\right]-Q(r_*)H=-\omega^2\Lambda(r_*)H, \label{H-SL}
\end{eqnarray}
with
\begin{eqnarray}
P(r_*)\!\!\!\!\!&=&\!\!\!\!\!e^{\nu}(\rho+p)r^2, \label{P-SL}\nonumber\\
Q(r_*)\!\!\!\!\!&=&\!\!\!\!\!2e^{2\nu}(\rho+p)\left[n+1-2\pi r^2(\rho+p)(3+{1}/{C_s^2})\right], \label{Q-SL}\nonumber\\
\Lambda(r_*)\!\!\!\!\!&=&\!\!\!\!\!e^{\nu}(\rho+p)r^2/{C_s^2}. \label{Lambda-SL}\nonumber
\end{eqnarray}
There are a series of real eigenfrequencies $\{\sigma_{n}\}$ and the corresponding normalized eigenfunctions $\{H_{n}(r_*)\}$, which are defined by the boundary conditions derived above, namely
(i) $H$ is regular at $r_*=0$, (ii) $(dH/dr_*)/H=R^2\omega^2/M$ at $r=R_*$. Following straightforwardly from the standard theory of Sturm-Liouville system and the fact that
$\Lambda(r=0)=0=\Lambda(r=R)$, the normalized eigenfunctions $\{H_{n}(r_*)\}$ form a complete orthogonal set obeying the completeness and orthogonality relationships \citep[see, e.g.,][]{Morse-2008}:
\begin{eqnarray}
\!\!\!\!\!&&\!\!\!\!\!\sum_{n}\Lambda(r_*)H_n(r_*)H_m(r_*')=\delta(r_*-r_*'); \label{H-Com} \\
\!\!\!\!\!&&\!\!\!\!\!\int_{0}^{R_*}\Lambda(r_*)H_n(r_*)H_m(r_*)dr_*=\delta_{n m}\label{H-Orth} .
\end{eqnarray}

\subsection{Numerical results and discussions}
To gauge the accuracy of the CA developed (or ICA as discussed
later)  in the present paper, we use EOS A \citep{model-A} to
construct a NS with compactness ${\cal C}=0.27$ and central
density $\rho_c=2.227\times 10^{-3} \textrm{km}^{-2}$ and evaluate
the QNMs of the star. Table \ref{Sigma-n} shows the frequencies of
the leading fluid modes obtained from exact calculation and CA
(\ref{Cowling-H}) developed above. We see that the lowest
eigenfrequency $\sigma_{0}$ of (\ref{H-SL}) provides us an
approximate value of the real part of $f$-mode frequency,
$\omega_{{\rm r},f}$, with a percentage error of around $20\%$.
For $n\geq 1$, the eigenfrequency $\sigma_{n}$ yields an accurate
approximation for the real part of the frequency of the $n$-th
$p$-mode, $\omega_{{\rm r},p_n}$. The percentage error is usually
much less than $1\%$ and decreases with increasing $n$ (see
Table~\ref{Sigma-n}). The improvement in accuracy from $f$-mode to
$p$-modes of the developed CA is expected. As the imaginary part
of eigenfrequency decreases from $f$-mode to $p$-modes, the
coupling between the spacetime and matter also gets weaker.

\begin{table*}
\centering \caption{Exact QNM frequency $(\omega_{\rm
r},\omega_{\rm i})$ and the CA frequencies $\sigma_n$ obtained
respectively from (\ref{Cowling-H}), (\ref{AAKS-H-decoupled}), MVS
\citep{MVSCowling}, Finn  \citep{Finn1988},  and LMI
\citep{Lindblom-1997} CA's, and (\ref{F-CA}) are compared for the
leading fluid modes of an EOS A star (${\cal C}=0.27$).}
\label{Sigma-n} \centering
\begin{tabular}{cccccccc}
\hline
 Mode      &        Exact       &  (\ref{Cowling-H})  &  (\ref{AAKS-H-decoupled}) & MVS      &        Finn       & LMI     &  (\ref{F-CA})  \\
               & ($\omega_{\rm r}$, $\omega_{\rm i}$)& $\sigma_n$ & $\sigma_n$ &   $\sigma_n$ &   $\sigma_n$ & $\sigma_n$ & $\sigma_n$\\
\hline
$f$               & (0.1447, 7.5E-5)   & 0.1163  & NA       & 0.1664 & 0.1773 & 0.1665 & NA        \\
$p_1$          & (0.3932, 3.0E-6)   & 0.3928  & 0.3934 & 0.4433 & 0.4806 & 0.3984 & 0.3903 \\
$p_2$          & (0.6130, 5.8E-7)   & 0.6135  & 0.6130 & 0.6489 & 0.7102 & 0.6159 & 0.6126  \\
$p_3$          & (0.7639, 5.3E-8)   & 0.7640  & 0.7639 & 0.7842 & 0.8525 & 0.7664 & 0.7636 \\
\hline
\end{tabular}
\end{table*}

\begin{figure}
\centering
\includegraphics[width=0.4\textwidth]{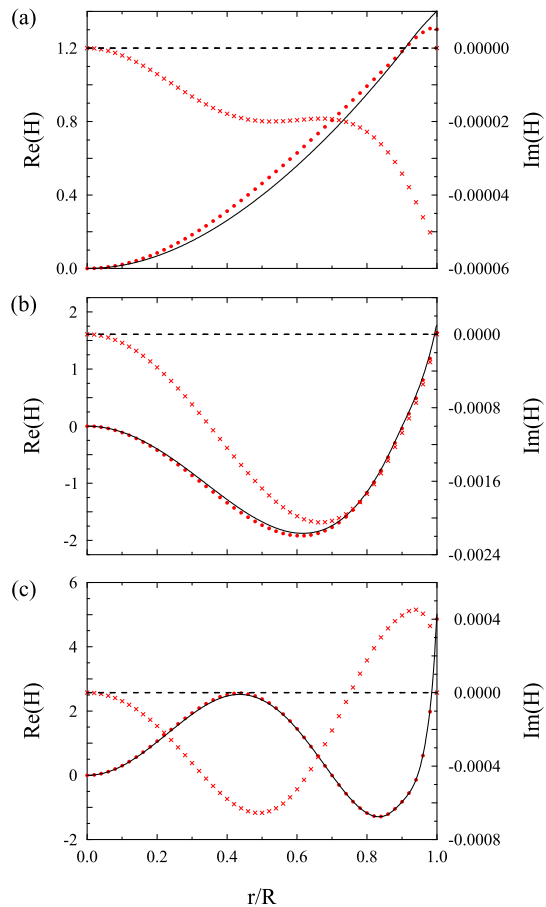}
\caption{The normalized $H$-functions for (a) the $f$-mode; (b)
the $ p_1$-mode; and (c) the $p_2$-mode of a star constructed with
EOS A ($\rho_c=2.227\times10^{-3}\,\texttt{km}^{-2}$, ${\cal
C}=0.27$) are plotted against the normalized radius $r/R$. The
solid and the dashed lines are respectively the real and the
imaginary parts of $H$-function under CA (\ref{Cowling-H}). The
dots and crosses respectively show the real and the imaginary
parts of the exact $H$-function. }\label{H-function}
\end{figure}

We also compare the $H$-function obtained from (\ref{Cowling-H})
and the exact $H$-function for the same star (EOS A, ${\cal
C}=0.27$). The comparison is shown in Fig.~\ref{H-function}. The
$H$-functions in developed CA and the exact AAKS formalism are
normalized so that
\begin{equation}\label{H-norm}
\int_{0}^{R_*}\Lambda(r_*)|H|^2dr_*=1\,. \nonumber
\end{equation}
Besides, we assume that the exact $H$-function is real at $r=R$.
The numerical results again confirm that the CA proposed here is
very accurate for $p$-modes. The discrepancy between the
approximate and the exact wave functions grows larger for
$f$-mode, which implies that the coupling between spacetime (i.e.,
$F$ and $S$) and matter is very weak for $p$-modes, but is not
negligible for $f$-mode.

\begin{figure}
\centering
\includegraphics[width=0.45\textwidth]{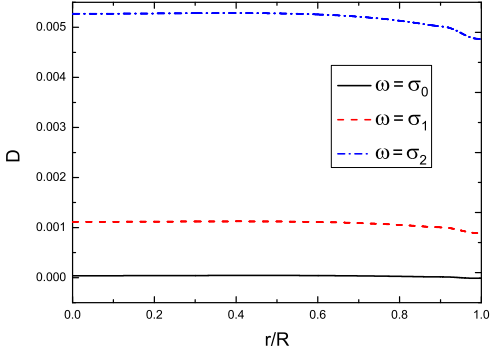}
\caption{The function $D(\omega,r)$ is plotted against $r/R$ with
$\omega=\sigma_0, \sigma_1, \sigma_2$.}\label{D-graph}
\end{figure}
Actually, as is mentioned before, the potentials $U_i$ and
couplings $\Delta_i$ in the exact $H$ equation
(\ref{AAKS-H-inside}) have a common denominator $D(\omega,r)$
given by (\ref{D}). It is clearly seen that the accuracy of the CA
(\ref{Cowling-H}), would decrease if $D(\omega,r)$ is small. As is
shown in Fig.~\ref{D-graph}, where $D$ is plotted against $r/R$
for different values of $\omega$ for the same star. $D$ decreases
with decreasing $\omega$. In fact, for $\omega=\sigma_0$ (i.e.,
the $f$-mode frequency under CA), $D$ vanishes at $r/R=0.947$.
Therefore, the omission of the influence of spacetime (i.e., $F$
and $S$) on oscillations of matter field may lead to appreciable
errors at low frequencies. On the other hand, we observe that
there are exactly $n$ nodes (excluding the origin) in the wave
function $H_n$, in agreement with standard theory of the
Sturm-Liouville eigen-system \citep[see, e.g.,][]{zettl2005sturm}.

For the purpose of comparison, approximate eigenfrequencies
obtained from (\ref{AAKS-H-decoupled}) are also listed in Tables
\ref{Sigma-n}, \ref{CA-SLy} and \ref{CA-Poly}. While Eq.
(\ref{AAKS-H-decoupled}) in general can yield almost exact
numerical results for $p$-modes because it has properly taken the
influence of $F$ on $H$ into consideration, it completely fails to
locate approximate position of the $f$-mode. As mentioned above,
the inability of (\ref{AAKS-H-decoupled}) to handle the $f$-mode
can be understood as a direct consequence of the fact that $D$
could vanish at certain positions inside the star at frequencies
close to the $f$-mode frequency.

\begin{table}
\begin{center}
  \caption{Exact complex eigenfrequencies $(\omega_{\rm r},\omega_{\rm i})$ and approximate real eigenfrequencies  $\sigma_n$ obtained
  respectively from CA
  (\ref{Cowling-H}), (\ref{AAKS-H-decoupled})  and (\ref{F-CA}) are listed for the leading fluid modes of  NSs
  constructed with the Sly EOS \citep{DH2001,anaEOS}  for
   different compactness ${\cal C}$. Entries labelled by `NA' signify
modes that cannot be located by a specific
approximation.}\label{CA-SLy}
\begin{tabular}{cccccc}
\hline \hline
${\cal C}$      & Mode & Exact   & (\ref{Cowling-H})   & (\ref{AAKS-H-decoupled})  & (\ref{F-CA})   \\
&& $(\omega_{\rm r},\omega_{\rm i})$ &  $\sigma_n$& $\sigma_n$&
$\sigma_n$\\\hline
 0.117  & $f$    & (0.0488, 1.162E-5)  & 0.0410 & NA     & NA     \\
 0.117  & $p_1$   & (0.1654, 4.866E-7) & 0.1651 & 0.1654 & 0.1628 \\
 0.117  & $p_2$   & (0.2079, 2.029E-8)  & 0.2079 & 0.2079 & 0.2071 \\
 0.117  & $p_3$   & (0.2390, 5.709E-9) & 0.2390 & 0.2390 & 0.2383 \\
 \hline
 0.210 & $f$    & (0.1047, 5.290E-5)  & 0.0848 & NA     & NA     \\
 0.210 & $p_1$   & (0.3281, 2.371E-6)  & 0.3279 & 0.3283 & 0.3245 \\
 0.210 & $p_2$   & (0.4995, 3.435E-8) & 0.4996 & 0.4995 & 0.4985 \\
 0.210 & $p_3$   & (0.6154, 6.028E-9)  & 0.6155 & 0.6154 & 0.6150 \\
\hline
 0.276  & $f$    & (0.1476, 7.570E-5)  & 0.1177 & NA     & NA     \\
 0.276  & $p_1$   & (0.4074, 1.884E-6)  & 0.4072 & 0.4075 & 0.4053 \\
 0.276  & $p_2$   & (0.6236, 6.806E-7)  & 0.6241 & 0.6236 & 0.6234 \\
 0.276  & $p_3$   & (0.8192, 7.455E-8)  & 0.8195 & 0.8192 & 0.8193 \\
 \hline
\end{tabular}
\end{center}
\end{table}

\begin{table}
\begin{center}
  \caption{Exact complex eigenfrequencies $(\omega_{\rm r},\omega_{\rm i})$ and approximate real eigenfrequencies  $\sigma_n$ obtained
  respectively from CA
  (\ref{Cowling-H}), (\ref{AAKS-H-decoupled})  and (\ref{F-CA}) are listed for the leading fluid modes of polytropic stars
 (EOS:  $p=100\rho^{1.8}$) with
different compactness ${\cal C}$. Entries labelled by `NA' signify
modes that cannot be located by a specific
approximation.}\label{CA-Poly}
\begin{tabular}{cccccc}
\hline \hline
${\cal C}$      & Mode & Exact   & (\ref{Cowling-H})   & (\ref{AAKS-H-decoupled})  & (\ref{F-CA})   \\
&& $(\omega_{\rm r},\omega_{\rm i})$ & $\sigma_n$& $\sigma_n$&
$\sigma_n$\\\hline
0.117 & $f$     & (0.0525, 1.329E-5)  & 0.0475 & NA     & NA     \\
0.117 & $p_1$   & (0.1160, 3.283E-6)  & 0.1149 & 0.1164 & 0.1074 \\
0.117 & $p_2$   & (0.1761, 2.680E-7)  & 0.1757 & 0.1762 & 0.1727 \\
0.117 & $p_3$   & (0.2338, 1.473E-8)  & 0.2336 & 0.2338 & 0.2320 \\
\hline
0.204 & $f$     & (0.1160, 5.771E-5)  & 0.1021 & NA     & NA     \\
0.204 & $p_1$   & (0.2337, 1.728E-5)  & 0.2307 & 0.2343 & 0.2202 \\
0.204 & $p_2$   & (0.3477, 5.490E-7)  & 0.3470 & 0.3478 & 0.3429 \\
0.204 & $p_3$   & (0.4584, 3.069E-8)  & 0.4582 & 0.4584 & 0.4561 \\
\hline
0.271 & $f$     & (0.1659, 5.795E-5)  & 0.1473 & NA     & NA     \\
0.271 & $p_1$   & (0.3012, 1.341E-5)  & 0.2948 & 0.3020 & 0.2925 \\
0.271 & $p_2$   & (0.4249, 2.832E-5)  & 0.4242 & 0.4250 & 0.4249 \\
0.271 & $p_3$   & (0.5496, 1.359E-5)  & 0.5498 & 0.5495 & 0.5501 \\
\hline

\end{tabular}
\end{center}
\end{table}

\subsection{Comparison with other CA schemes}
We compare the CA developed here with other existing schemes,
namely, MVS \citep{MVSCowling}, Finn  \citep{Finn1988}, and LMI
\citep{Lindblom-1997}, and the $F$-$S$ CA (\ref{F-CA}) to be
developed in Appendix \ref{SI-C}. The numerical results obtained
from these approximations are shown in Table~\ref{Sigma-n}. It is
clear that all the three CA's developed in this paper can provide
more accurate results for $p$-modes in comparison with other
schemes proposed previously. While Eq.~(\ref{AAKS-H-decoupled})
yields the best  accuracy for $p$-modes, Eq.~(\ref{Cowling-H})
successfully handles both $f$ and $p$-modes.

The physical origin of the high accuracy of the CA proposed here
can be understood as follows. In the differential line element
given by (\ref{polar-metric-perturbation}), there are three
unknown metric coefficients, $H_0$, $H_1$ and $K$, which are to be
determined by solving the perturbed Einstein equation. In most, if
not all, CA's established previously  the approximation $H_0=K=0$
was always assumed. In the Newtonian limit, such an approximation
amounts to neglecting the change in gravitational potential in
stellar oscillations \citep{Cowling}, which is not a good
approximation for $f$-mode and low-order $p$-modes. In contrast,
in the present paper we only take the approximation $S=0$, i.e.,
$H_0=K$, which is strictly valid in the Newtonian limit. Hence, we
expect that our CA can provide much improved accuracy. In fact,
Table~\ref{Sigma-n} and additional comparisons in
Tables~\ref{CA-SLy} and \ref{CA-Poly}, also show that the high
accuracy of the CA proposed here is generic and independent of the
EOS and the compactness of the NS.

On the other hand, as an aside, we find that the CA's of
(\ref{Cowling-H}), MVS and Finn can be  unified by a single
second-order differential equation in $H$:
\begin{eqnarray}
\frac{d^2 H}{dr_*^2}\!\!\!\!\!&+&\!\!\!\!\! \left[V_{H1}(r)-\frac{e^{(\lambda-\nu)/2}}{\alpha}\frac{d\alpha}{dr}\right]
\frac{d H}{dr_*}+\alpha\left[\frac{\omega^2}{C_s^2}- V_{H0}(r) \right.\nonumber\\
\!\!\!\!\!&+&\!\!\!\!\!\left. 4 \pi  (\beta-1)
e^\nu(\rho+p)(3+\frac{1}{C_s^2})\right]H=0,\label{all-CA}
\end{eqnarray}
where the expressions for $\alpha$ and $\beta$ are listed in
Table~\ref{Cowcon} for different approximations. Indeed, the
improved accuracy of the current CA is closely tied to the choice
$\beta=1$, which originates from the non-vanishing  effects due to
$H_0$ and $K$.

\begin{table}
\centering \caption{Definition of $\alpha$ and $\beta$ in
different forms of CA.} \label{Cowcon}
\begin{tabular}{ccc}
\hline
\hspace{5 mm}CA\hspace{5 mm}             & $\hspace{5 mm}\alpha\hspace{5 mm}$                     & $\hspace{5 mm}\beta\hspace{5 mm}$ \\
\hline
(\ref{Cowling-H})  & 1                           & 1       \\
MVS           & 1                                    & 0       \\
Finn           & $1-\frac{16\pi r^2(\rho+p)}{l(l+1)}$ & 0       \\
\hline
\end{tabular}
\end{table}

\section{Inverse-Cowling Approximation for Spacetime Modes}\label{ICA-w-mode}

\subsection{Spacetime equation}
As shown by \citet{Wu-2007}, in polar $w$-mode oscillations, the
perturbation  in pressure is small  and accordingly the
contribution of the $H$ and $dH/dr_*$ terms in
(\ref{AAKS-S-inside}) is negligible. Hence, polar $w$-modes can be
described by (\ref{AAKS-S-decoupled}), forming the ICA proposed in
the present paper. To our knowledge, this is the first time that a
single second-order  equation for polar $w$-modes has been
proposed and verified numerically (see the following discussion).

It is interesting to note that the ICA equation
(\ref{AAKS-S-decoupled}) can be further recast into a simple
second-order equation:
\begin{eqnarray} \frac{d^2 \widetilde{S}}{d
r_*^2}+[\omega^2-\widetilde{V}(\omega, r)]\widetilde{S}=0,
\label{AAKS-ICA}
\end{eqnarray}
where
\begin{eqnarray}
\!\!\!\!&&\!\!\!\!\widetilde{S}(r_*)=\exp\left[\frac{1}{2}\int_0^{r_*}U_{S1}(\omega,r')dr_*'\right]S(r_*),\label{transformation}\\
\!\!\!\!&&\!\!\!\!\widetilde{V}(\omega,
r)=\!V_{S0}\!+\!U_{S0}\!+\!\frac{1}{2}\frac{dU_{S1}}{dr_*}\!+\!\frac{U_{S1}^2}{4}.\label{Vs-tilde-in}
\end{eqnarray}
$\widetilde{S}$ is proportional to $S$ as $r\rightarrow \infty$.
The above equation bears a strong resemblance to the Regge-Wheeler
(RW) equation governing the propagation of axial GWs
\citep{Chandrasekhar1}, save for the frequency dependence of the
potential $\widetilde{V}(\omega, r_*)$, which, physically
speaking, can be considered as the result of dispersion.
\subsection{Numerical results and discussions}
The spacetime $w$-modes can be located by solving (\ref{AAKS-S-decoupled}) or
equivalently (\ref{AAKS-ICA}) with the outgoing boundary condition. The numerical results of ICA proposed here (the columns labelled with ICA) are compared with the exact polar $w$-mode
eigenfrequencies in Table~\ref{ICA-polar-w-mode}, where the exact and approximate QNM frequencies of the leading twelve polar $w$-modes (including two $w_{\rm II}$ modes) for
the above-mentioned EOS A NS (${\cal C}=0.27$) are listed. The agreement between the exact frequencies and the ICA frequencies is almost perfect, especially for $\omega_{\rm r}$ and higher order modes.
\begin{table*}
\begin{center}
\caption{QNM frequencies obtained from various ICA  schemes for the leading polar $w$-modes of a EOS A star (${\cal C}=0.27$). The first two modes are classified as the $w_{\rm II}$ modes. Entries labelled by `NA' signify modes that cannot be located by a specific approximation.}\label{ICA-polar-w-mode}
\begin{tabular}{cccccccc}
\hline
~Exact~  & ~Exact~  & ~~ICA~~ & ~~ICA~~ & HFICA    & HFICA    & LMI & LMI \\
 $\omega_{\rm r}$&$\omega_{\rm i}$&$\omega_{\rm
r}$&$\omega_{\rm i}$&$\omega_{\rm r}$&$\omega_{\rm
i}$&$\omega_{\rm r}$&$\omega_{\rm i}$
\\
\hline
0.2067 & 0.7935 & 0.2065 & 0.7932 & 0.3304 & 0.8922 & NA       & NA       \\
0.4593 & 0.3891 & 0.4599 & 0.3884 & 0.5076 & 0.4980 & NA       & NA       \\
0.5096 & 0.1777 & 0.5095 & 0.1781 & 0.5102 & 0.1131 & 0.4348   &
0.0605   \\
0.8730 & 0.3025 & 0.8730 & 0.3028 & 0.8335 & 0.2781 & 0.7924   &
0.1244   \\
1.2266 & 0.3507 & 1.2266 & 0.3508 & 1.2006 & 0.3416 & 1.1446   &
0.1555   \\
1.5782 & 0.3837 & 1.5782 & 0.3838 & 1.5589 & 0.3788 & 1.4961   &
0.1749   \\
1.9302 & 0.4093 & 1.9301 & 0.4093 & 1.9148 & 0.4062 & 1.8476   &
0.1892   \\
2.2828 & 0.4308 & 2.2828 & 0.4308 & 2.2700 & 0.4287 & 2.2004   &
0.2013   \\
2.6359 & 0.4500 & 2.6359 & 0.4500 & 2.6250 & 0.4485 & 2.5537   &
0.2122   \\
2.9894 & 0.4682 & 2.9894 & 0.4683 & 2.9799 & 0.4670 & 2.9079   &
0.2231   \\
3.3429 & 0.4860 & 3.3428 & 0.4860 & 3.3344 & 0.4850 & 3.2619   &
0.2347   \\
3.6960 & 0.5038 & 3.6960 & 0.5038 & 3.6884 & 0.5030 & 3.6158   &
0.2468   \\
\hline
\end{tabular}
\end{center}
\end{table*}

\subsection{High-frequency ICA}
As mentioned above, both $U_{S0}(\omega,r)$ and $U_{S1}(\omega,r)$
go as $1/\omega^2$ at high frequencies. Thus, we expect that
$w$-mode QNMs with high frequencies can be approximately located
by omitting terms $U_{S0}(\omega,r)$ and $U_{S1}(\omega,r)$ in
(\ref{AAKS-S-decoupled}), namely,
\begin{eqnarray}
\frac{d^2 S}{d r_*^2}+[\omega^2-V_{S0}(r)]S=0. \label{HFA}
\end{eqnarray}
Such a scheme is termed as the high-frequency inverse-Cowling
approximation (HFICA) in the present paper. As shown in
Table~\ref{ICA-polar-w-mode}, HFICA works well in the high
frequency regime. In fact, HFICA can successfully explain the
high-frequency asymptotic behavior of polar $w$-modes of compact
stars  \citep{Zhang_asym}.

\section{Conclusion and discussion}
In this paper,  we propose an alternative description for polar
oscillations of compact stars, which consists of two coupled
second-order ODEs in  one metric variable ($S$) and one matter
variable ($H$). The $H$-$S$ approach developed here readily leads
to CA (ICA) with unprecedented accuracies for polar $p$-modes
($w$-modes) when the two second-order ODEs are decoupled. Besides,
under the CA the wave functions $H_n$ ($n=0,1,2,\ldots$) of the
fluid modes (including the $f$-mode and the $p$-modes) are shown
to form a complete orthonormal set, which paves the way for
further developing perturbation schemes to improve the accuracies
of the CA by including the effect of GW radiation damping.

We note that there are other theories for pulsations of compact
stars which also result in two coupled second-order ODEs. For
example, in the LMI formalism \citep{Lindblom-1997}  the metric
variable $H_0(r, t)$ and the fluid variable $\delta U(r, t)\equiv
\delta p(r, t)/(\rho+p)+H_0(r, t)/2$ are used to formulate a
relativistic theory for stellar pulsations. Similar to the
development of the present paper, the two coupled second-order
ODEs in the LMI formalism can be decoupled by adopting the
approximations $H_0=0$ and $\delta U=0$, leading to the so called
LMI CA and LMI ICA, respectively. The LMI CA was applied to find
the frequency of the $f$-mode of a polytropic NS with
$M=1.4M_\odot$ and the percentage error was about $20\%$ for the
case $l=2$ \citep{Lindblom-1997}, which is similar to ours.
Besides, as shown in Table~\ref{Sigma-n}, the LMI CA  can yield
quite accurate frequencies for the $p$-modes. On the other hand,
the frequencies of the leading polar $w$-modes obtained from the
LMI ICA  are listed in the Table~\ref{ICA-polar-w-mode}. It is
clear, from the numerical results, that the performance of the LMI
ICA is not that satisfactory.

It is essential to note that the success of the CA and ICA
established in the present paper relies on a judicious selection
of the two independent variables, namely $H$ and $S$. While the
former is proportional to the change in the Eulerian pressure,
i.e., a proper Newtonian quantity, the latter vanishes in the weak
field limit. Thus, the interplay between $H$ and $S$ in fluid
modes and spacetime modes is expected to be small, which is
numerically verified in the above discussion. According to our
experience, the choice of the independent variables $H$ and $S$
(especially the latter) is the most crucial factor leading  to the
high precision of the ICA scheme developed here. We note that
several other schemes for CA \citep[see, e.g.,][and references
therein]{Lindblom-1990} and ICA \citep{Andersson-1996} based on
the LD formalism have been proposed. Yet, the accuracies of these
schemes are worse than that achieved in the present paper. This
clearly demonstrates the importance of the choice of variables in
carrying out the decoupling scheme.

On the other hand, the mutual influences between spacetime and
matter,  which are neglected in CA and ICA developed in the
present paper, can also be included in a perturbative manner. As
supported by the accuracies of these two approximations (see
numerical data shown in the paper), we expect that relevant
perturbation series could converge rapidly to yield satisfactory
results. For example, GWs emitted in $p$-modes and hence the
finite lifetime of these QNM could be evaluated perturbatively.
Research along such direction is now underway and relevant results
will be published  elsewhere in due course.

Under ICA developed here, the equation of motion for $\tilde{S}$
in (\ref{AAKS-ICA}), resembles the RW equation governing the
motion of axial perturbations of compact stars. Such resemblance
suggests that  axial and polar perturbations in compact stars
could be studied in a parallel way. In fact, it has recently been
shown that polar $w$-mode QNMs can be inferred from axial $w$-mode
QNMs asymptotically \citep{Zhang_asym}. The formulation
established here is likely to shed light on the related studies.

In a series of papers, \citet{Tsui05:3}, \citet{Tsui05:2} and
\citet{Tsui:prd} developed an inversion scheme to infer the
internal structure and the EOS of a NS from a few of its leading
axial QNMs. Despite the success in the axial case, the inversion
method cannot be directly applied to the polar type oscillations
due to the complicated form of the equations of motion, which
arises from the interplay between matter and spactime. The CA and
ICA proposed in our paper, where both fluid modes and spactime
modes are governed by second-order ODEs analogous to the RW
equation, could pave the way for the generalization of the
inversion scheme to include polar QNMs.

Lastly, as $F$ is proportional to the change in the Newtonian
potential in the weak field limit, it can also reveal the motion
of matter indirectly and replace the role of $H$. In fact, one can
also formulate another equivalent description for stellar
pulsation using the variables $F$ and $S$ to develop another
two-variable approach, namely the $F$-$S$ approach  (see
Appendix~\ref{SI-C} for detailed discussion). Similar to the case
in the LMI formalism mentioned above, such an $F$-$S$ approach can
yield valid CA with accuracies worse than those obtained from
(\ref{AAKS-H-decoupled}) and (\ref{Cowling-H}), but fails to
predict accurate polar $w$-modes. The ICA in the $F$-$S$ approach
is identical to the HFICA in the $H$-$S$ approach and hence works
well only in the high frequency limit (see Appendix~\ref{SI-C} and
Table~\ref{ICA-polar-w-mode}). In  contrast,
Eq.~(\ref{AAKS-S-decoupled}), or equivalently (\ref{AAKS-ICA}),
used in the $H$-$S$ approach can locate all polar $w$-modes
including the low-frequency $w_{\rm II}$ modes with excellent
numerical accuracies. In conclusion, the $H$-$S$ approach proposed
here successfully leads to feasible and accurate CA and ICA
schemes in a unified framework and, in this regard, it outperforms
other two-variable approaches to the best of our knowledge.

\section*{Acknowledgments}
This work is supported in part by the Hong Kong Research Grants
Council (Grant No: 401807) and the direct grant (Project ID:
2060330) from the Chinese University of Hong Kong. We thank
L.M.~Lin, H.K.~Lau, and P.O.~Chan for helpful discussions.

\newcommand{\noopsort}[1]{} \newcommand{\printfirst}[2]{#1}
\newcommand{\singleletter}[1]{#1} \newcommand{\switchargs}[2]{#2#1}

\begin{onecolumn}

\appendix
\newpage
\section{AAKS Formalism}\label{SI-A}

In the AAKS formalism \citep{Allen-1998}, polar QNM oscillation of
compact stars with a frequency $\omega$  are governed by the
following three coupled second-order differential equations:
\begin{eqnarray}
& & \omega^2 S+\frac{d^2 S}{d r_*^2}+\frac{2e^{\nu}}{r^3}\left[ 2\pi r^3(\rho+3p)+m-(n+1)r\right]S = -\frac{4e^{2\nu}}{r^5}\left[ \frac{(m+4\pi p r^3)^2}{r-2m}+4\pi \rho r^3-3m\right]F,\label{Allen-S-w}\\
& & \omega^2 F+\frac{d^2F}{d r_*^2}+\frac{2e^{\nu}}{r^3}\left[ 2\pi r^3(3\rho+p)+m-(n+1)r\right]F = -2\left[ 4\pi r^2(p+\rho)-e^{-\lambda}\right]S+8\pi(\rho+p)re^{\nu}( 1-\frac{1}{C_s^2})H,\label{Allen-F-w}\\
& & \frac{\omega^2}{C_s^2} H+\frac{d^2H}{d r_*^2}+\frac{e^{(\nu+\lambda)/2}}{r^2}\frac{d H}{d r_*}\left[ (m+4\pi pr^3)(1-\frac{1}{C_s^2})+2(r-2m)\right] +\frac{2e^{\nu}}{r^2}\left[2\pi r^2(\rho+p)(3+\frac{1}{C_s^2})-(n+1)\right]H \nonumber \\
& &= (m+4\pi pr^3)(1-\frac{1}{C_s^2})\frac{e^{(\lambda-\nu)/2}}{2r}
\left(\frac{e^{\nu}}{r^2}\frac{d F}{d r_*}-\frac{d S}{d
r_*}\right)+\left[\frac{(m+4\pi pr^3)^2}{r^2(r-2m)}(1+\frac{1}{C_s^2})-\frac{m+4\pi pr^3}{2r^2}( 1-\frac{1}{C_s^2})-4\pi r(3p+\rho)\Bigg]S \right. \nonumber \\
& & +\frac{e^{\nu}}{r^2}\left[\frac{2(m+4\pi pr^3)^2}{r^2(r-2m)}\frac{1}{C_s^2}-\frac{m+4\pi pr^3}{2r^2}(1-\frac{1}{C_s^2})-4\pi r(3p+\rho)\right]F, \label{Allen-H-w}
\end{eqnarray}
In addition to the above three equations, $S(r)$, $F(r)$ and
$H(r)$ should also satisfy the Hamiltonian constraint:
\begin{eqnarray}
&&\frac{d^2F}{d r_*^2}-\frac{e^{(\nu+\lambda)/2}}{r^2}(m+4\pi r^3p)\frac{d F}{d r_*}+\frac{e^{\nu}}{r^3}[12\pi r^3\rho-m-2(n+1)r]F \nonumber \\
&-& re^{-(\nu+\lambda)/2}\frac{d S}{d r_*}+\left[ 8\pi
r^2(p+\rho)-(n+3)+\frac{4m}{r}\right]S +\frac{8\pi
r}{C_s^2}e^{\nu}(\rho+p)H = 0.\label{H-constraint-w}
\end{eqnarray}
By combining (\ref{Allen-F-w}) and the Hamiltonian constraint (\ref{H-constraint-w}) together, a simple expression for $H$ can be derived:
\begin{eqnarray}
8\pi(\rho+p) r e^{\nu} H = \frac{e^{(\lambda+\nu)/2}}{r^2}(m+4\pi pr^3)\frac{dF}{dr_*}
+\omega^2 F+\frac{e^{\nu}}{r^3}(3m+4\pi pr^3)F+{e^{-(\nu+\lambda)/2}}{r}\frac{dS}{dr_*}+(n+1)S,
\label{H-expression}
\end{eqnarray}
which is often used to calculate the value of $H$, instead of
integrating (\ref{Allen-H-w}), and termed as the modified
Hamiltonian constraint hereafter.


As for the region outside the star the fluid variable $H$ vanishes
identically, the two metric variables $S$ and $F$ evolve as:
\begin{eqnarray}
& & \omega^2 S+\frac{d^2 S}{d r_*^2}+\frac{2(r-2M)}{r^4}\left[M-(n+1)r\right]S
    = -\frac{4 M}{r^7}(r-2 M)(7 M - 3 r)F, \label{Allen-S-out}\\
& & \omega^2
F+\frac{d^2F}{dr_*^2}+\frac{2(r-2M)}{r^4}\left[M-(n+1)r\right]F
    = 2(1-\frac{2M}{r})S, \label{Allen-F-out}
\end{eqnarray}
which can be obtained from (\ref{Allen-S-w}) and (\ref{Allen-F-w})
by setting $p=\rho=0$ and $m(r)=M$.

\section{Derivation of $H$-$S$ scheme}\label{SI-B}

In the following, we eliminate the variable $F$ from the equations
of motion shown in Appendix~\ref{SI-A} and establish the so called
 $H$-$S$ approach, which leads to accurate CA for the fluid modes
(especially the $p$-modes) and ICA for the $w$-modes.

 First of
all, for the region $r<R$, we rewrite the modified Hamiltonian
constraint (\ref{H-expression}) and its derivative with respect to
$r_*$ as:
\begin{eqnarray}
8\pi(\rho+p) e^{\nu} r H &=& A_{F1}\frac{dF}{dr_*}+A_{F0}F+A_{S1}\frac{dS}{dr_*}+A_{S0}S, \label{H-1}\\ 8\pi(\rho+p)e^{\nu} \frac{d}{d r_*}(r H) &= &B_{F1}\frac{dF}{dr_*}+B_{F0}F+B_{S1}\frac{dS}{dr_*}+B_{S0}S, \label{H-2}
\end{eqnarray}
where
\begin{eqnarray}
A_{S1} & = & {e^{-(\nu+\lambda)/2}}{r},                    \label{As1}\\
A_{S0} & = & n+1,                                          \label{As0}\\
A_{F1} & = & \frac{e^{(\lambda+\nu)/2}}{r^2}(m+4\pi pr^3), \label{Af1}\\
A_{F0} & = & \omega^2+\frac{e^{\nu}}{r^3}(3m+4\pi
pr^3),\label{Af0}\\
B_{S1} & = & 1+n+e^{-\lambda}-4 \pi r^2 (p+\rho),\label{Bs1}\\
B_{S0} & = & \frac{e^{(\nu-\lambda)/2}}{r}\left[2(1+n)-e^{-\nu} r^2 \omega^2-4 \pi r^2 (p+\rho) \left(1+r \nu'\right)\right],\label{Bs0}\\
B_{F1} & = & \omega ^2+e^{\nu}\left[\frac{m}{r^3}+4 \pi (2 p+\rho)\right],\label{Bf1}\\
B_{F0} & = & \frac{e^{(\nu -\lambda )/2}}{r^4} \left\{e^{\nu}(3m-4
\pi  r^3 \rho)-
             \frac{r^2 \nu'}{2}\left[r^2 \omega^2-2 e^{\nu} (1+n-8 \pi  r^2 (p+\rho))\right]\right\},\label{Bf0}
\end{eqnarray}
with $\nu'=d\nu/dr$. To arrive at (\ref{H-2}), Eqs.~(\ref{Allen-S-w}) and (\ref{Allen-F-w}) have been used to eliminate $d^2 F/dr^2_*$ and $d^2 S/dr^2_*$.

From (\ref{H-1}) and (\ref{H-2}), we can express $F$ and $dF/dr_*$ in terms of $S$, $dS/dr_*$, $H$ and $dH/dr_*$:
\begin{eqnarray}
F&=&\alpha_{S1}\frac{dS}{dr_*}+\alpha_{S0}S+\alpha_{H1}\frac{dH}{dr_*}+\alpha_{H0}H,\label{F-in}\\
\frac{d F}{d r_*}&=&\beta_{S1}\frac{dS}{dr_*}+\beta_{S0}S+\beta_{H1}\frac{d H}{dr_*}+\beta_{H0}H,
\label{DF-in}
\end{eqnarray}
where
\begin{eqnarray}
\alpha_{S1} & = & -\frac{1}{D}\bigg\{e^{-(\lambda +\nu )/2} r\omega ^2+\frac{e^{(\nu -\lambda )/2}}{r^2} \left[4 \pi  r^3 (p+\rho )-e^{\lambda }(m+4 \pi  r^3 p)(1+n)+4 \pi r^2 e^{\lambda }(m+4 \pi r^3 p)(p+\rho )\right]\bigg\},\label{alpha-s1}\\
\alpha_{S0}& = & -\frac{1}{D}\bigg\{\left[n+1+\frac{m}{r}+4\pi r^2 p\right]\omega^2+\frac{e^{\nu}}{r^3}\left[-(1+n)(m-4 \pi r^3 \rho)+4 \pi r^2 e^{\lambda}(m+4 \pi r^3 p)(1+8\pi r^2 p)(p+\rho)\right]\bigg\},\label{alpha-s0}\\
\alpha_{H1}& = & -\frac{8\pi e^{(\lambda+3\nu)/2}(\rho+p)}{Dr}(m+4\pi pr^3),\label{alpha-h1}\\
\alpha_{H0}& = & \frac{8\pi e^{\nu}(\rho+p)r}{D}\left[\omega^2+4 \pi e^{\nu} (p+\rho)\right],\label{alpha-h0}
\end{eqnarray}
\begin{eqnarray}
\beta_{S1}& = & -\frac{1}{D}\bigg\{\left[n+2-\frac{m}{r}-4 \pi r^2 \rho \right]\omega^2+\frac{e^{\nu}}{r^3}\left\{m \left[1+n-4 \pi r^2 (p+\rho)\right]+4 \pi r^3 \left[\rho -n p+12 \pi  r^2 p(p+\rho)\right]\right\}\bigg\},\label{beta-s1}\\
\beta_{S0}& = &-\frac{e^{-(\lambda+\nu)/2}}{D}\bigg\{-r\omega^4+\frac{e^{\nu}}{r^2}\left\{e^{\lambda} (m+4 \pi r^3 p)\left[1+n-8 \pi r^2 (p+\rho)\right]+2r(n+1)-3m-4\pi r^3(2p+\rho)\right\}\omega^2 \nonumber\\&   & +\frac{e^{2\nu}}{r^4}\left\{m (1+n+\frac{2m}{r})\right.+2(m+4 \pi r^3 p)
                 \left[\frac{m}{r}+n-e^{\lambda}(1+n)^2\right]+4\pi r(\rho+p)(m+4\pi r^3 p) \nonumber\\
           &   & \left.\left\{r-2 e^{\lambda}\left[m-r(1+2n)+4\pi r^3 p\right]\right\}+4 \pi r^3 \rho
                 [1+n-\frac{2m}{r}]\right\}\bigg\}   ,\label{beta-s0}\\
\beta_{H1}
           & = & \frac{8\pi e^{\nu}(\rho+p) r}{D} \left[\omega^2+\frac{e^{\nu}}{r^3}(3m+4\pi pr^3)\right],
                                                                            \label{beta-h1}\\
\beta_{H0}
           & = & \frac{8\pi e^{(3\nu -\lambda)/2}(\rho+p)}{D}\bigg\{
                 \left[1+\frac{e^{\lambda}}{r}(m+4 \pi r^3 p)\right]\omega^2
                 +\frac{e^{\nu}}{r^3}\left\{4 \pi r^3 (p+\rho)-2 e^{\lambda}(m+4 \pi r^3 p)
                 \left[1+n-8 \pi r^2(p+\rho)\right]\right\}\bigg\},\label{beta-h0}
\end{eqnarray}
and $D$ is defined in (\ref{D}).

For $r>R$, the stellar pressure $p$ and density $\rho$ vanish and
so does $H$, then $F$ in (\ref{F-in}) reduces to the following
form:
\begin{eqnarray}
F=\alpha_{S1}^{(\rm e)}\frac{dS}{dr_*}+\alpha_{S0}^{(\rm e)}S, \label{F-out}
\end{eqnarray}
which is a linear combination of $S$ and $dS/dr_*$. The
coefficients $\alpha_{S0}^{(\rm e)}$ and $\alpha_{S1}^{(\rm e)}$
can be readily obtained from $\alpha_{S0}$ and $\alpha_{S1}$ by
taking $\rho=p=0$ and $m=M$:
\begin{eqnarray}
\alpha_{S1}^{(\rm e)} & = & -\frac{1}{D^{(\rm e)}}\left[r\omega^2-\frac{1}{r^2}M(1+n)\right],  \label{alpha-s1-out}\\
\alpha_{S0}^{(\rm e)} & = & -\frac{1}{D^{(\rm
e)}}\left[(n+1+\frac{1}{r}M)\omega^2-\frac{1}{r^4}M(1+n)(r-2M)\right],
\label{alpha-s0-out}
\end{eqnarray}
with $D^{(\rm e)}$ being:
\begin{eqnarray}
D^{(\rm e)} & = & \omega^4+\frac{1}{r^4}M(-7 M+4
r)\omega^2-\frac{2}{r^7}M^2(1+n)(r-2 M).
\end{eqnarray}

With the expressions of $F$ and  $d F/d r_*$ derived above, we
manage to find (\ref{AAKS-S-inside}) and (\ref{AAKS-H-inside}) for
the fields $H$ and $S$
 inside the
star. The expressions for potential $U_i$ and coupling $\Delta_i$
in these two equations are as follows:
\begin{eqnarray}
U_{S1}(\omega, r)  & = & -V^S_{F0}(r) \alpha_{S1}(\omega, r) , \\
U_{S0}(\omega, r)  & = & V^S_{F0}(r) \alpha_{S0}(\omega, r) , \\
U_{H1}(\omega, r)  & = & -V^H_{F1}(r) \beta_{H1}(\omega, r) -V^H_{F0}(r) \alpha_{H1}(\omega, r) , \\
U_{H0}(\omega, r)  & = & V^H_{F1}(r) \beta_{H0}(\omega, r) +V^H_{F0}(r) \alpha_{H0}(\omega, r) , \\
\Delta_{S1}(\omega, r)  & = & V^H_{S1}(r) + V^H_{F1}(r) \beta_{S1}(\omega, r)  + V^H_{F0}(r) \alpha_{S1}(\omega, r) , \\
\Delta_{S0}(\omega, r)  & = & V^H_{S0}(r) + V^H_{F1}(r) \beta_{S0}(\omega, r)  + V^H_{F0}(r) \alpha_{S0}(\omega, r) , \\
\Delta_{H1}(\omega, r)  & = & V^S_{F0}(r) \alpha_{H1}(\omega, r) , \\
\Delta_{H0}(\omega, r)  & = & V^S_{F0}(r) \alpha_{H0}(\omega, r) ,
\end{eqnarray}
where $\alpha_i$ and $\beta_i$  are defined in
(\ref{alpha-s1})-(\ref{beta-h0}), and
\begin{eqnarray}
V^S_{F0} & = & -\frac{4e^{2\nu}}{r^5}\left[ \frac{(m+4\pi p r^3)^2}{r-2m}+4\pi \rho r^3-3m\right], \label{Vs-f0}\\
V^H_{F1} & = & (m+4\pi pr^3)(1-\frac{1}{C_s^2})\frac{e^{(\lambda+\nu)/2}}{2r^3}, \label{Vh-f1}\\
V^H_{F0} & = & \frac{e^{\nu}}{r^2}\left[\frac{2(m+4\pi pr^3)^2}{r^2(r-2m)}\frac{1}{C_s^2}-\frac{m+4\pi pr^3}{2r^2}(1-\frac{1}{C_s^2})-4\pi r(3p+\rho)\right], \label{Vh-f0} \\
V^H_{S1} & = & -(m+4\pi pr^3)(1-\frac{1}{C_s^2})\frac{e^{(\lambda-\nu)/2}}{2r}, \label{Vh-s1} \\
V^H_{S0} & = & \frac{(m+4\pi
pr^3)^2}{r^2(r-2m)}(1+\frac{1}{C_s^2})-\frac{m+4\pi
pr^3}{2r^2}(1-\frac{1}{C_s^2})-4\pi r(3p+\rho). \label{Vh-s0}
\end{eqnarray}

Since the functions $\alpha_{H1}(\omega, r)$ and
$\alpha_{H0}(\omega, r)$ both contain a $(\rho+p)$ factor, they
vanish outside the star. Therefore, $\Delta_{H0}(\omega, r)$ and
$\Delta_{H0}(\omega, r)$ vanish outside the star. The system
reduces into one second-order ODE (\ref{AAKS-S-out}) in $S$
outside the star, where
\begin{eqnarray}
U_{S1}(\omega, r)  & = & -V^{S(\rm e)}_{F0}(r) \alpha^{(\rm e)}_{S1}(\omega, r), \\
U_{S0}(\omega, r)  & = & V^{S(\rm e)}_{F0}(r) \alpha^{(\rm e)}_{S0}(\omega, r),
\end{eqnarray}
with the $\alpha^{(\rm e)}$ coefficients defined in (\ref{alpha-s1-out}) and (\ref{alpha-s0-out}), and
\begin{eqnarray}
V^{S(\rm e)}_{S0} & = & \frac{2}{r^4}(r-2 M)[(1+n)r-M], \label{Vs-s0-out}\\
V^{S(\rm e)}_{F0} & = & \frac{4}{r^7}(r-2 M) M (3 r-7 M).\label{Vs-f0-out}
\end{eqnarray}

\section{$F$-$S$ Approach} \label{SI-C}
\subsection{Formalism}
As both $F$ and $H$ could reveal the Newtonian aspect of stellar
pulsations, it is natural to expect that an alternative approach
using $F$ and $S$ to study stellar pulsations is possible. In
fact, it is straightforward to make use of the Hamiltonian
constraint (\ref{H-constraint-w}) to eliminate the matter field
$H(r_*)$ in (\ref{Allen-F-w}) \citep{Allen-1998}:
\begin{eqnarray}
& &\frac{\omega^2 F}{C_s^2}+\frac{d^2F}{d r_*^2}-
\left(1-\frac{1}{C_s^2}\right)\frac{e^{(\nu+\lambda)/2}}{r^2}(m+4\pi
pr^3)\frac{d F}{d r_*}+\frac{e^{\nu}}{r^3}\left[4\pi
r^3\left(3\rho+\frac{p}{C_s^2}\right)-m\left(1-\frac{3}{C_s^2}\right)-2(n+1)r\right]F
\nonumber \\ &&=
\left(1-\frac{1}{C_s^2}\right)re^{-(\nu+\lambda)/2}\frac{d S}{d
r_*} +\left[
2e^{-\lambda}+\left(1-\frac{1}{C_s^2}\right)(n+1)-8\pi(p+\rho)r^2\right]S.
\label{Allen-F2}
\end{eqnarray}
(\ref{Allen-S-w}) and (\ref{Allen-F2}) together form two coupled
second-order equations in $F$ and $S$, which can be used to locate
QNMs of compact stars and lead to the so-called {\it F-S} approach
in the present paper.

\subsection{Cowling approximation}
 Similar to the $H$-$S$ approach,  the $S$ terms in
(\ref{Allen-F2}) are expected to be negligible for fluid mode
pulsations, yielding
\begin{eqnarray}
\frac{\omega^2 F}{C_s^2}+\frac{d^2F}{d r_*^2}-
\left(1-\frac{1}{C_s^2}\right)\frac{e^{(\nu+\lambda)/2}}{r^2}(m+4\pi
pr^3)\frac{d F}{d r_*} +\frac{e^{\nu}}{r^3}\left[4\pi
r^3\left(3\rho+\frac{p}{C_s^2}\right)-m\left(1-\frac{3}{C_s^2}\right)-2(n+1)r\right]F
=0.  \label{F-CA}
\end{eqnarray}
When subjected to suitable boundary condition (as discussed in the
following), (\ref{F-CA}) can give good approximation to $p$-mode
pulsations of realistic compact stars, thereby resulting in the CA
for $p$-modes in the {\it F-S} approach.
 As usual, the regularity boundary condition $F \propto
r_*^l$ holds around the origin. As the fluid variable $H$ is
bounded, it can be shown that the correct boundary condition near
the stellar surface is $F \sim (R_*-r_*)^{N+1}$. Upon imposing the
two boundary conditions mentioned above on the solution of
(\ref{F-CA}), we can find the approximate values of the real part
of the eigenfrequencies of $p$-modes, which are listed and
compared with the exact values $\omega_{\rm r}$ in
Tables~\ref{Sigma-n}, \ref{CA-SLy} and
 \ref{CA-Poly}. For $p$-modes of NSs constructed with realistic EOSs, e.g., A and SLy (see Tables~\ref{Sigma-n} and \ref{CA-SLy}), the percentage error in $\omega_{\rm
r}$ is usually less than a few percent and decreases with the mode
order and the compactness of the star. However, the accuracy of
the CA scheme (\ref{F-CA}) could substantially worsen for
polytropic stars (especially those with large polytropic indices)
in spite of the fact that it still improves with increasing
compactness and mode order (see Table~\ref{CA-Poly}). In
comparison with the two CA schemes proposed in the $H$-$S$
approach, i.e., (\ref{Cowling-H}) and (\ref{AAKS-H-decoupled}),
(\ref{F-CA}) is in general the least accurate one. Moreover,  we
cannot locate the $f$-mode oscillation under the CA in  {\it F-S}
approach.

 Starting from CA (\ref{F-CA}) in the $F$-$S$ approach, we can also get a
standard Sturm-Liouville equation in terms of the variable $G
\equiv F/p$,
\begin{eqnarray}
\frac{d}{dr_*}\left[P_G(r_*)\frac{dG}{dr_*}\right]-Q_G(r_*)G+\omega^2\Lambda_G(r_*)G=0,\label{F2-Eq}
\end{eqnarray}
with
\begin{eqnarray}
P_G(r_*)&=&\frac{p^2}{e^{\nu}(\rho+p)},\\
Q_G(r_*) &=&-\frac{p}{2r(r-2m)}\left[16\pi
mr(\rho+3p)-64\pi^2r^4p^2-8\pi
r^2(\rho+p)+1\right]+\frac{(r-2m)p}{2r^3}\nonumber\\
&&-\frac{p^2}{r^3(\rho+p)}\left[4\pi
r^3(3\rho+\frac{p}{C_s^2})-m(1-\frac{3}{C_s^2})-2(n+1)r\right],\\
\Lambda_G(r_*)&=&\frac{p^2}{e^{\nu}(\rho+p)C_s^2}.
\end{eqnarray}
We note that  the coefficients $P_G(r_*)$ and $\Lambda_G(r_*)$ are
strictly positive, as required  for a normal Sturm-Liouville
equation \citep[see, e.g.,][]{zettl2005sturm}. Besides,
$\Lambda_G$ vanishes at both $r=0$ and $r=R$.  Following directly
from the physical boundary condition on $F$ as mentioned above,
$G$ tends to a finite constant at $r_*=R_*$ and
\begin{eqnarray}
\left[\frac{1}{G}\frac{dG}{dr_*}\right]_{r_*=R_*}=\frac{2(4N+1)M-(5N+2)R}{(N+2)R^2}-\frac{N
R^2\omega^2}{(N+2)M}.
\end{eqnarray}
Thus, a standard Sturm-Liouville eigenvalue system is established
and the normalized eigenfunctions of (\ref{F2-Eq}), $G_n(r_*)$,
where $n=1,2,3,\ldots$, form a complete set and satisfy the
orthogonality relation \citep[see, e.g.,][]{zettl2005sturm}:
\begin{eqnarray}
 \int_0^{R_*}\frac{p^2
G_m(r_*)G_n(r_*)}{e^\nu(\rho+p)C_s^2}dr_*=\delta_{mn}.
\end{eqnarray}


As mentioned above, in the $F$-$S$ CA scheme (\ref{F-CA}) we fail
to locate the $f$-mode. Hence, the $n$-th eigenfunction $G_n$
($n=1,2,3,\ldots$) in fact corresponds to the $n$-th $p$-mode. In
Fig.~\ref{G} we show the scaled eigenfunction $\tilde{G}_n$
defined as:
\begin{equation}\label{scaledG}
 \tilde{G}_n (r_*) \equiv \frac{p
G_n(r_*)}{\sqrt{e^\nu(\rho+p)C_s^2}},
\end{equation}
which satisfies the orthonormal condition $\int_0^{R_*}
\tilde{G}_m(r_*)\tilde{G}_n(r_*)dr_*=\delta_{mn}$. There are $n-1$
nodes (excluding the two endpoints $r=0$ and $r=R$) in the wave
function $\tilde{G}_n$ (or equivalently $G_n)$. In particular, the
wave function of the $p_1$-mode is nodeless. This observation
 further confirms that the $p_1$-mode is indeed the ground state of
the eigenvalue equation  (\ref{F2-Eq}) \citep[see,
e.g.,][]{zettl2005sturm} and explains  why the $f$-mode is missing
in the CA scheme (\ref{F-CA}) of the $F$-$S$ approach.

\begin{figure}
\centering
\includegraphics[height=8.0cm]{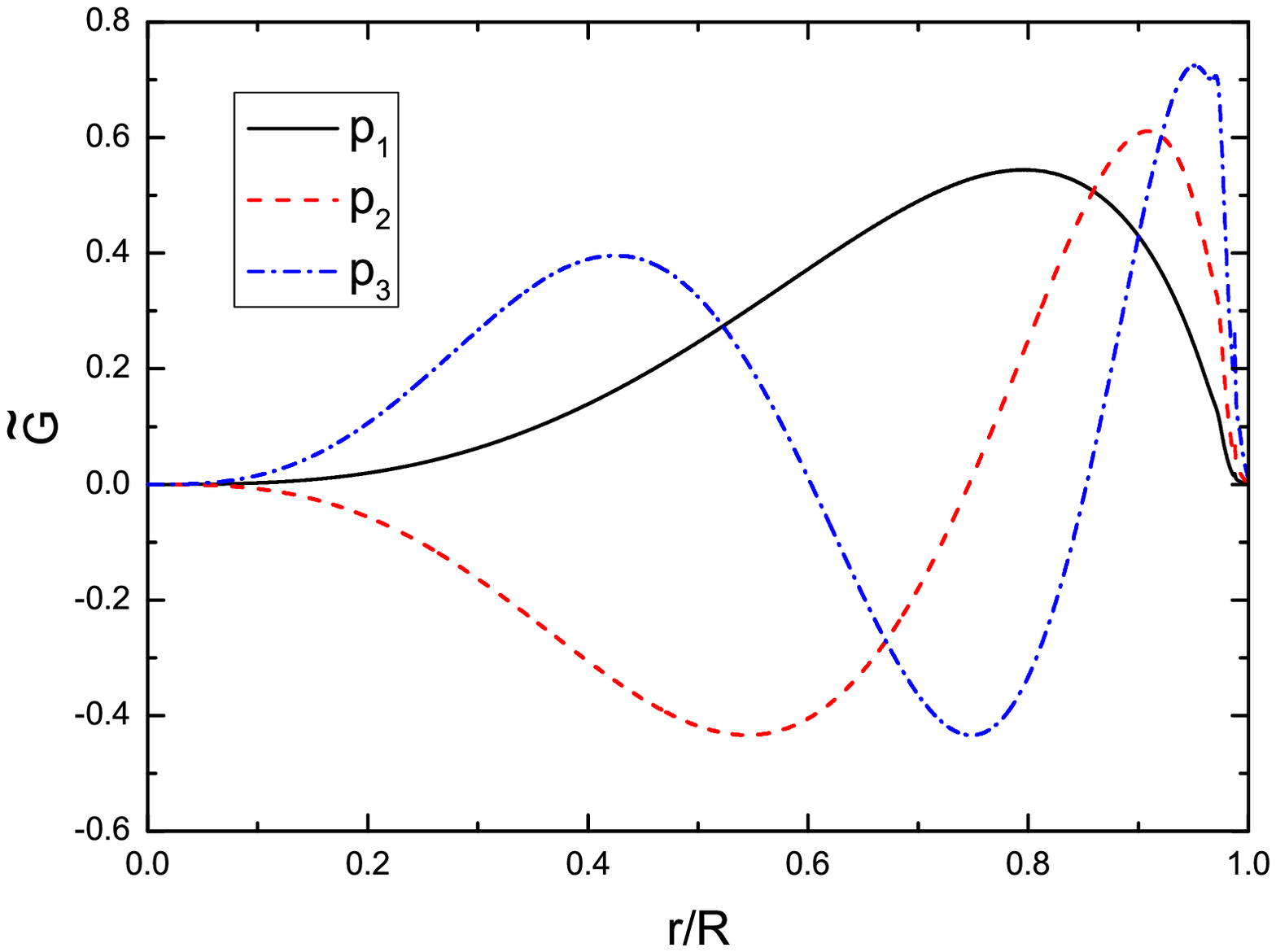}
\caption{The scaled function $\tilde{G}_n$ of a  NS with  SLy EOS
\citep{DH2001,anaEOS}  and compactness ${\cal C}=0.276$ is plotted
against $r/R$ for $n=1$ (solid line), 2 (dashed line), 3
(dot-dashed line), which correspond to $p_1$, $p_2$ and $p_3$
modes, respectively.}\label{G}
\end{figure}

\subsection{Inverse-Cowling approximation}
We can also obtain the ICA for polar $w$-modes within the {\it
F-S} approach by neglecting the $F$ term in (\ref{Allen-S-w}),
i.e.,
\begin{eqnarray}
\omega^2 S+\frac{d^2 S}{d r_*^2}+\frac{2e^{\nu}}{r^3}\left[ 2\pi
r^3(\rho+3p)+m-(n+1)r\right]S=0. \label{F-ICA}
\end{eqnarray}
Interestingly, such an ICA
 is merely the HFICA in the $H$-$S$ approach, which works well only for $w$-modes with high
frequencies  (see Table~\ref{ICA-polar-w-mode} for the accuracies
and validity of HFICA in the $H$-$S$ approach). It reflects the
fact that the metric variable $F$ also participates in the
dynamics of stellar pulsations and its contribution to
(\ref{Allen-S-w}) is non-negligible for $w$-modes with low
frequencies.

\end{onecolumn}

\bsp

\label{lastpage}

\end{document}